\documentclass[preprints,article,accept,moreauthors,pdftex]{Definitions/mdpi} 
\firstpage{1} 
\pubvolume{1}
\issuenum{1}
\articlenumber{0}
\pubyear{2021}
\copyrightyear{2020}
\datereceived{} 
\dateaccepted{} 
\datepublished{} 
\hreflink{https://doi.org/} % If needed use \linebreak
\pdfoutput=1

\usepackage{xspace}
\newcommand*{\sqs}{\ensuremath{\sqrt{s}}\xspace}
\newcommand*{\pT}{\ensuremath{p_\mathrm{T}}\xspace}

\newcommand*{\kT}{\ensuremath{k_\mathrm{T}}\xspace}
\newcommand*{\pTjet}{\ensuremath{p_\mathrm{T}^{\mathrm{jet}}}\xspace}
\newcommand*{\GeV}{\ensuremath{\mathrm{GeV}}\xspace}
\newcommand*{\TeV}{\ensuremath{\mathrm{TeV}}\xspace}

\Title{The partonic origin of multiplicity scaling in heavy and light flavor jets}

\TitleCitation{The partonic origin of multiplicity scaling}

\Author{Zolt\'an Varga $^{1,2}$*\orcidA{} and R\'obert V\'ertesi $^{1}$\orcidB{}}

\AuthorNames{Zoltán Varga, Róbert Vértesi}

\AuthorCitation{Varga, Z.; Vértesi, R.}

\address{%
$^{1}$ \quad Wigner Research Centre for Physics, P.O. Box 49, H-1525 Budapest, Hungary\\
$^{2}$ \quad Budapest University of Technology and Economics, M\H{u}egyetem rkp. 3., H-1111 Budapest, Hungary}

\corres{Correspondence: varga.zoltan@wigner.hu}

\abstract{It has recently been shown that a KNO-like scaling is fulfilled inside the jets, which indicates that the KNO scaling is violated by complex vacuum-QCD processes outside the jet development, such as single and double parton scattering or softer multiple-parton interactions. In the current work we investigated the scaling properties of heavy-flavor jets using Monte-Carlo simulations. We found that while jets from leading-order flavor-creation processes exhibit a flavor-dependent pattern, heavy-flavor jets from production in the parton shower follow the inclusive-jet pattern. This suggests that the KNO-like scaling is driven by initial hard parton production and not by processes in the later stages of the reaction.}

\keyword{High-energy hadron collisions; jets; KNO scaling; heavy-flavor; dead-cone effect} 

\begin{document}
\end{paracol}
%%%%%%%%%%%%%%%%%%%%%%%%%%%%%%%%%%%%%%%%%%

\section{Introduction}

Final-state multiplicities in small colliding systems are known to follow a negative binomial distribution (NBD) regardless of the collision species over several orders of magnitude of energy ranges~\cite{Chew:1985qsa,Giovannini:1985mz,ALICE:2017pcy}. It has also been observed in $e^+e^-$ collisions that the multiplicity distributions at different collision energies collapse into a single distribution when the so-called Koba--Nielsen--Olesen (KNO) scaling~\cite{Koba:1972ng,Polyakov:1970lyy} is applied. The KNO-scaling was, however, found to be violated at higher energies and in more complex, hadronic collision systems \cite{UA5:1985fid,EuropeanMuon:1987tnv}. The origin of the scaling, and the reason for its breakdown is still not understood completely, although  many explanations have been proposed in the past decades \cite{Lam:1983vw,Kudo:1985qfd,Burgers:1987za,Hegyi:1996wb,Matinyan:1998ja}.
In our earlier study we found that a KNO-like scaling is fulfilled within the jets. This indicates that KNO scaling is violated by complex quantum-chromodynamics (QCD) processes outside the jet development, such as single and double-parton scatterings or softer multiple-parton interactions (MPI)~\cite{Vertesi:2020utz}.

In our manuscript we investigate the scaling properties of heavy-flavor (HF) jets in comparison to an inclusive jet sample. Heavy flavor is mostly produced in hard (large momentum-exchange) processes, in the early stages of the collision. The most relevant perturbative QCD processes that contribute to the production cross section are leading-order (LO) flavor creation, and next-to-leading order (NLO) gluon splitting as well as flavor excitation~\cite{Ilten:2017rbd}. The parton shower and fragmentation of heavy-flavor jets is different from light-flavor jets due to two main reasons: the color charge effect, that is, heavy flavor jets are initiated by quarks as opposed to light-flavor jets that are mostly gluon-initiated~\cite{Apolinario:2022vzg}; and the dead-cone effect, meaning that small-angle gluon radiations off a massive parton are forbidden in QCD, and as a consequence, heavy-flavor fragmentation is harder and results in different jet substructures~\cite{Dokshitzer:1991fd,ALICE:2021aqk,Kucera:2021mfd,Varga:universe5050132}. 

In the current work we model both light and heavy-flavor jets using the PYTHIA 8 Monte-Carlo event generator~\cite{Sjostrand:2014zea}, and we differentiate the samples according to the process the jets are created in. Whether the KNO-like scaling observed for inclusive jets is retained or violated in heavy-flavor jets can shed light on the origin of the scaling itself, and also on possible mechanisms that are responsible for the violation of the scaling in heavy-flavor jets. The methods we present can further be used to gain insight to the flavor-dependent evolution of the jets. Future measurements targeted on the scaling of light and heavy-flavor jets can also serve as a validation tool for heavy-flavor production and fragmentation models.

%The structure of our paper is organized as follows. In Section \ref{sec:method} we describe the analysis method. In Section \ref{sec:results} we give a detailed description of our results. Finally, we summarize our results in Section \ref{sec:results}.

\section{Analysis Method}
\label{sec:method}

We simulated proton-proton collisions at \sqs$=7$ \TeV center-of-mass energy utilizing the PYTHIA 8 (version 8.226) Monte-Carlo event generator with the Monash tune and HardQCD settings~\cite{Sjostrand:2014zea,Corke_2011}.
PYTHIA 8 is tuned to describe both the fundamental physical observables of the leading hard process and the underlying event, and it is known to reproduce final-state multiplicities well \cite{CMS:2013ycn,CMS:2017eoq}. In PYTHIA 8 the hard parton scatterings and decays are simulated using LO matrix elements (ME). These are amended by initial and final state radiations, which create the parton shower (PS) in perturbative QCD calculations based on DGLAP splitting kernels~\cite{Sjostrand:2014zea}, as well as soft and hard multiple-parton interactions integrated into a single framework~\cite{Sjostrand:1987su}. The hadronic final state is then produced with the Lund string fragmentation model~\cite{Sjostrand:1982fn}.

Using the possibility in PYTHIA 8 to restrict event generation to certain hard processes, we created four different sets of data. As the baseline for our study, we used an inclusive-jet sample, where any hard QCD scattering process was allowed above an appropriately selected value of the minimal momentum transfer in the hardest process ($\hat{\pT}$) depending on the jet transverse momentum (\pTjet), as detailed in \cite{Varga:2018isd}. Next, we used samples with ME flavor creation, where hard $2\rightarrow 2$ parton scatterings were allowed only with heavy-flavor outgoing partons: $gg\rightarrow b\Bar{b}(c\Bar{c})$ and $q\Bar{q}\rightarrow b\Bar{b}(c\Bar{c})$. This provided wide-angle heavy-flavor jets created directly in the leading process of the event.
Finally, we created a sample that is dominated by b-jets from the PS, by allowing only those $2\rightarrow 2$ processes that do not directly create heavy flavor: $gg\rightarrow gg$, $gg\rightarrow q\Bar{q}$, $qg\rightarrow qg$, $q\Bar{q}\rightarrow gg$, $q\Bar{q}\rightarrow q'\Bar{q'}$ (where incoming HF is allowed, but only light flavor exits), and finally three more processes: $q\Bar{q'}\rightarrow qq'$, $q\Bar{q'}\rightarrow q\Bar{q'}$ and $\Bar{q}\Bar{q'}\rightarrow \Bar{q}\Bar{q'}$ (where outgoing and incoming flavors are the same and $q$ and $q'$ may be of the same flavor)~\cite{Giles:2015gluon}.
In this case the heavy quark pair is produced in a later step, e.g. in a $g\rightarrow b\Bar{b}$ gluon splitting process, typically with smaller opening angles. The heavy quarks then often manifest in the final state as secondary jets besides the leading jet, or may even end up in the same jet.

In all cases, charged-particle jets were clusterized from final-state charged pions, kaons and (anti)protons with \pT$>0.15$ \GeV using the anti-\kT jet-clustering algorithm~\cite{Cacciari2008} with a resolution parameter of $R=0.7$ in the mid-rapidity range $|\eta|<1$ and full azimuth coverage. The reconstructed jets were categorized in 20 different \pTjet ranges, from 15 \GeV up to 400 \GeV.
In the case of the charm and beauty jet samples, the corresponding heavy quark was required to fall within the cone of the selected jet, similarly to jet-tagging methods that are utilized in the experiment~\cite{CMS:2012feb,ALICE:2021wct}.

\section{Results and Discussion}
\label{sec:results}

\begin{figure}[ht!]
\centering
\includegraphics[width=0.5\columnwidth]{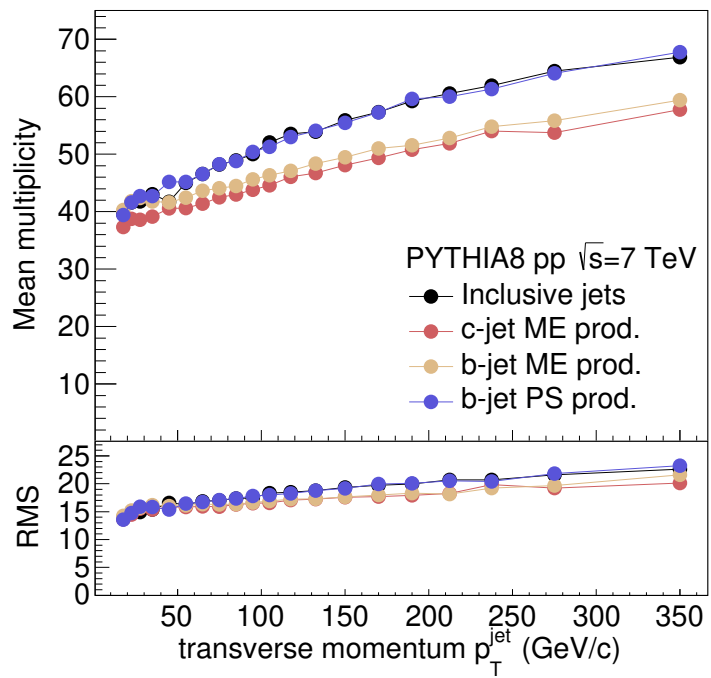}%
\caption{The mean (top panel) and RMS values (bottom panel) of the charged-hadron multiplicity distributions at $|\eta|<1$ for inclusive jets, for charm and beauty jets from ME-level production as well as for beauty jets from production in the PS, in function of \pTjet.}
\label{fig:multMeans}
\end{figure}

In Fig.~\ref{fig:multMeans} we plot the mean and RMS values of the event multiplicity ($N$) distributions at central pseudorapidity ($|\eta|<1$), in function of \pTjet, separately for inclusive jets, b-jets and c-jets from ME flavor creation as well as for b-jets from parton shower processes. As one expects, events having jets with a higher \pTjet contain more final-state hadrons on the average, and the distribution also gets broader toward higher \pTjet. Heavy-flavor jets from ME flavor creation correspond to a lower average multiplicity at a given \pTjet, while heavy-flavor from the parton shower follows the trend of inclusive jets. The difference is especially prominent for higher \pTjet.

\begin{figure}[ht!]
\centering
\includegraphics[width=0.4\columnwidth]{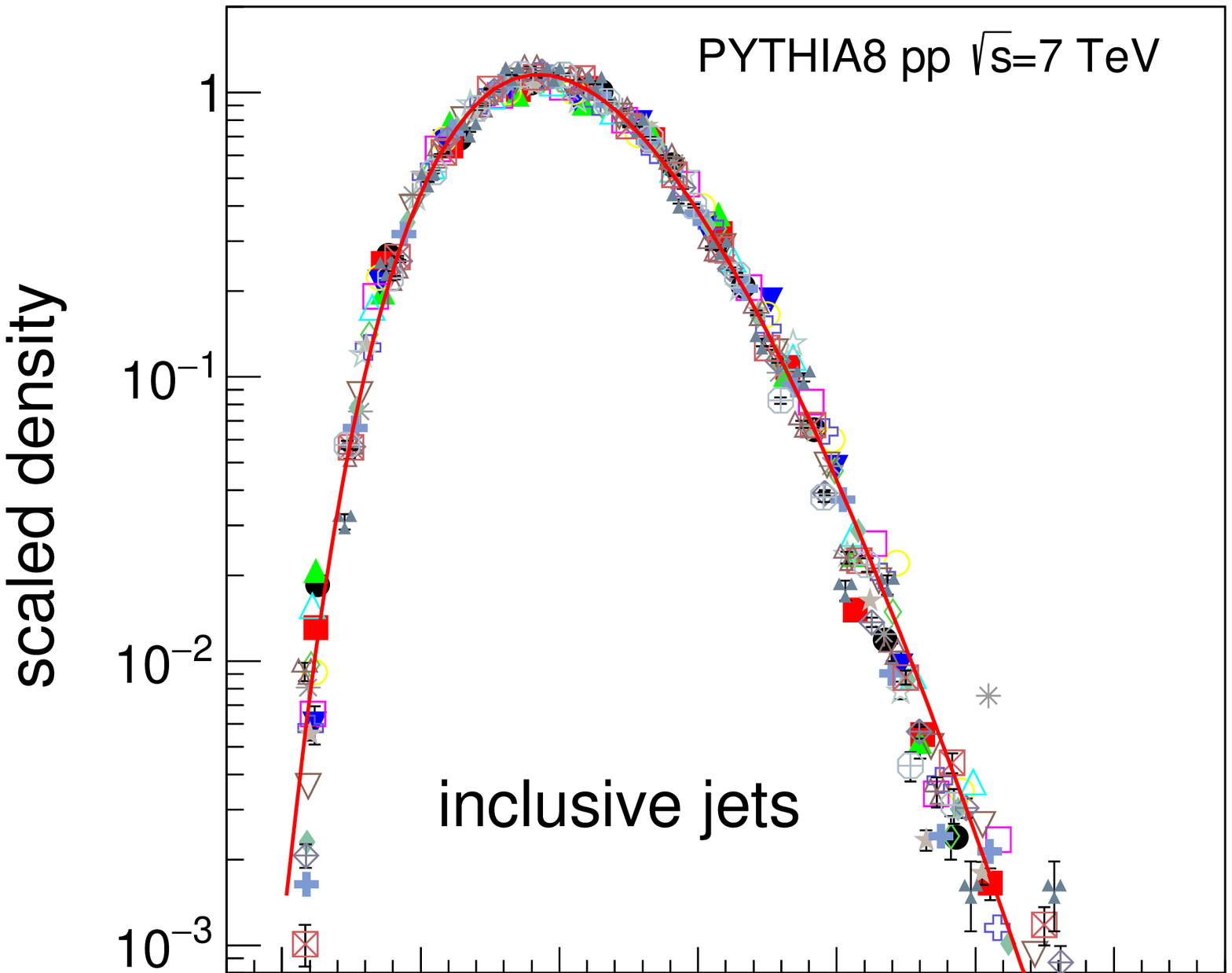}%
\includegraphics[width=0.4\columnwidth]{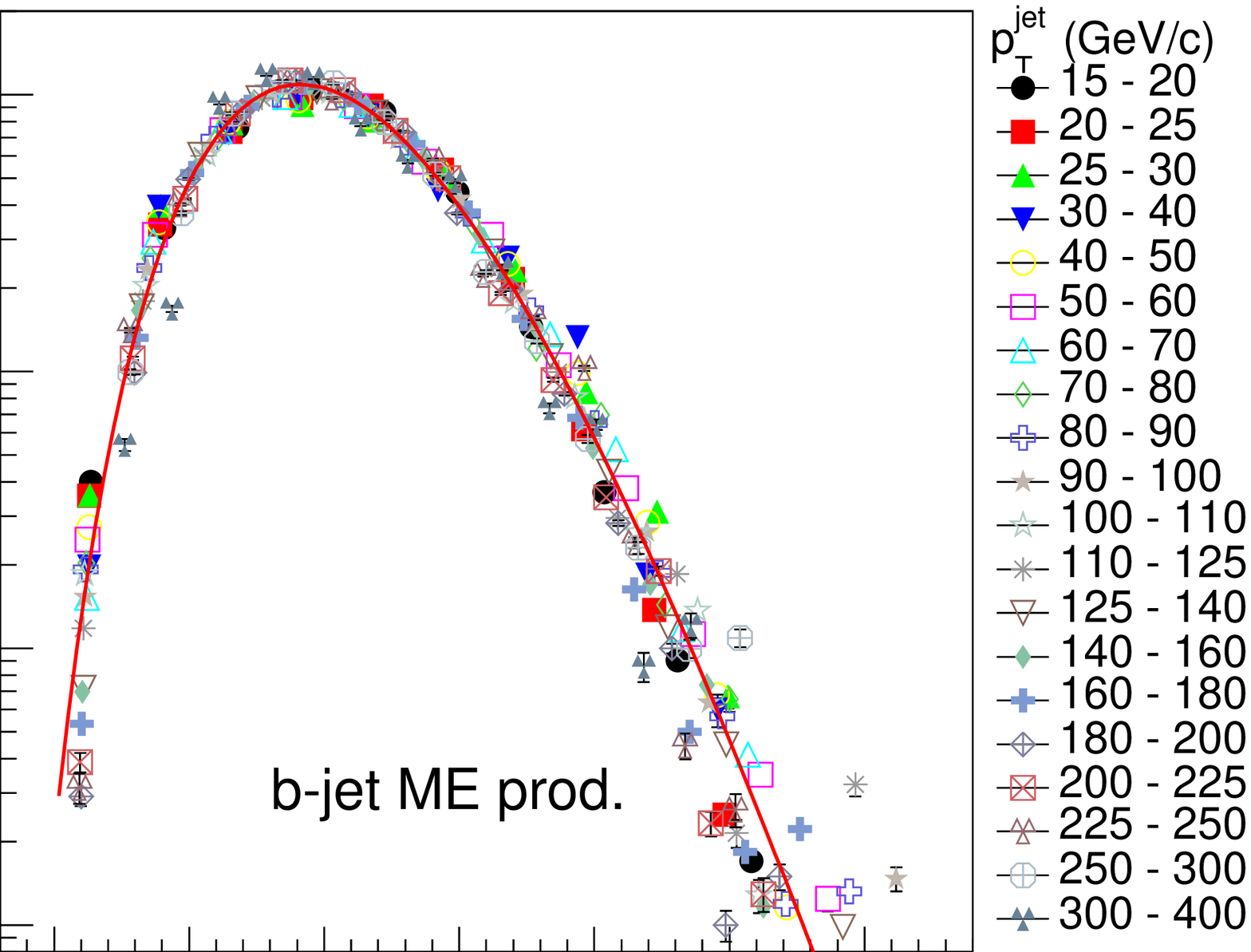}
\includegraphics[width=0.4\columnwidth]{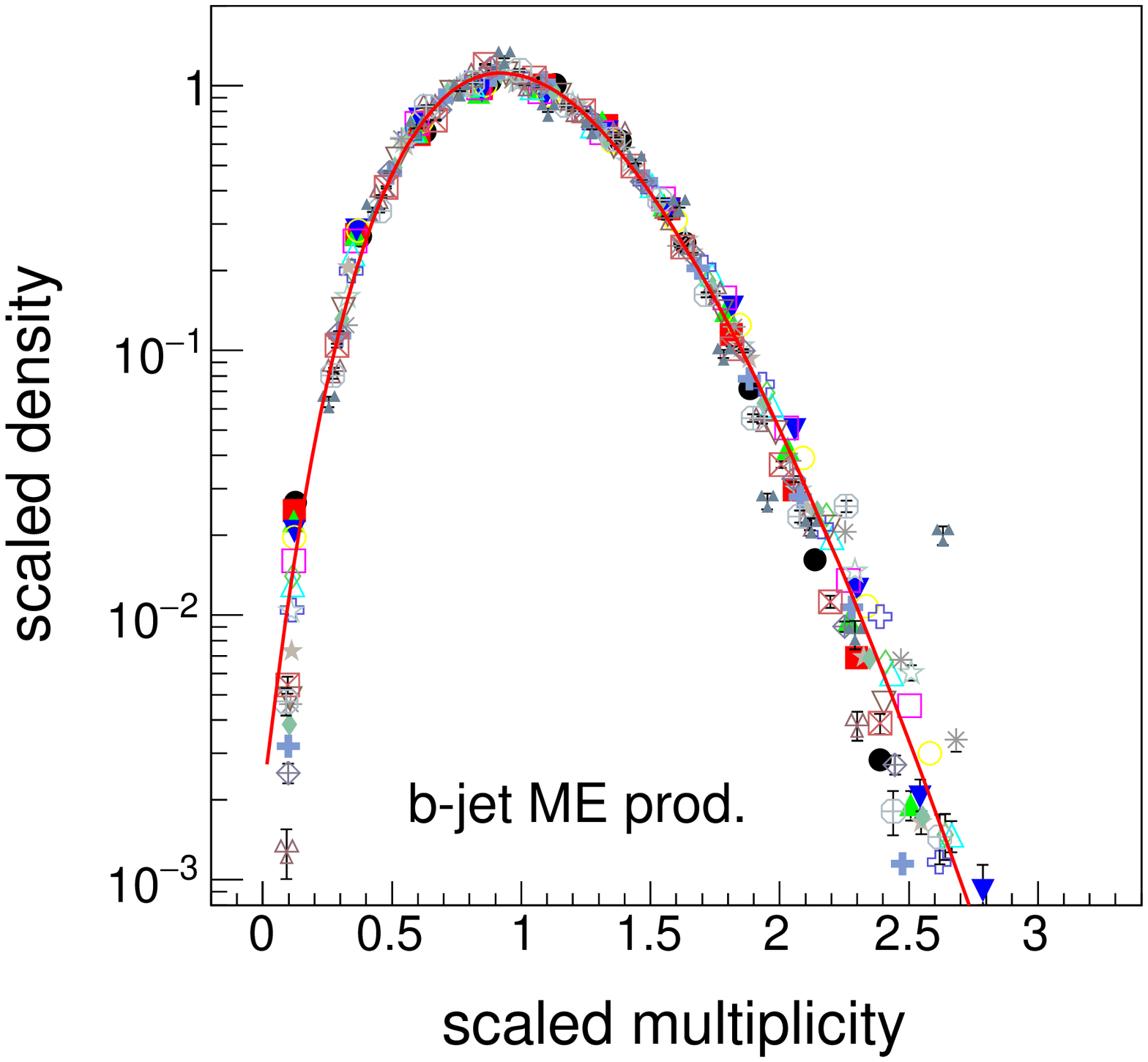}%
\includegraphics[width=0.4\columnwidth]{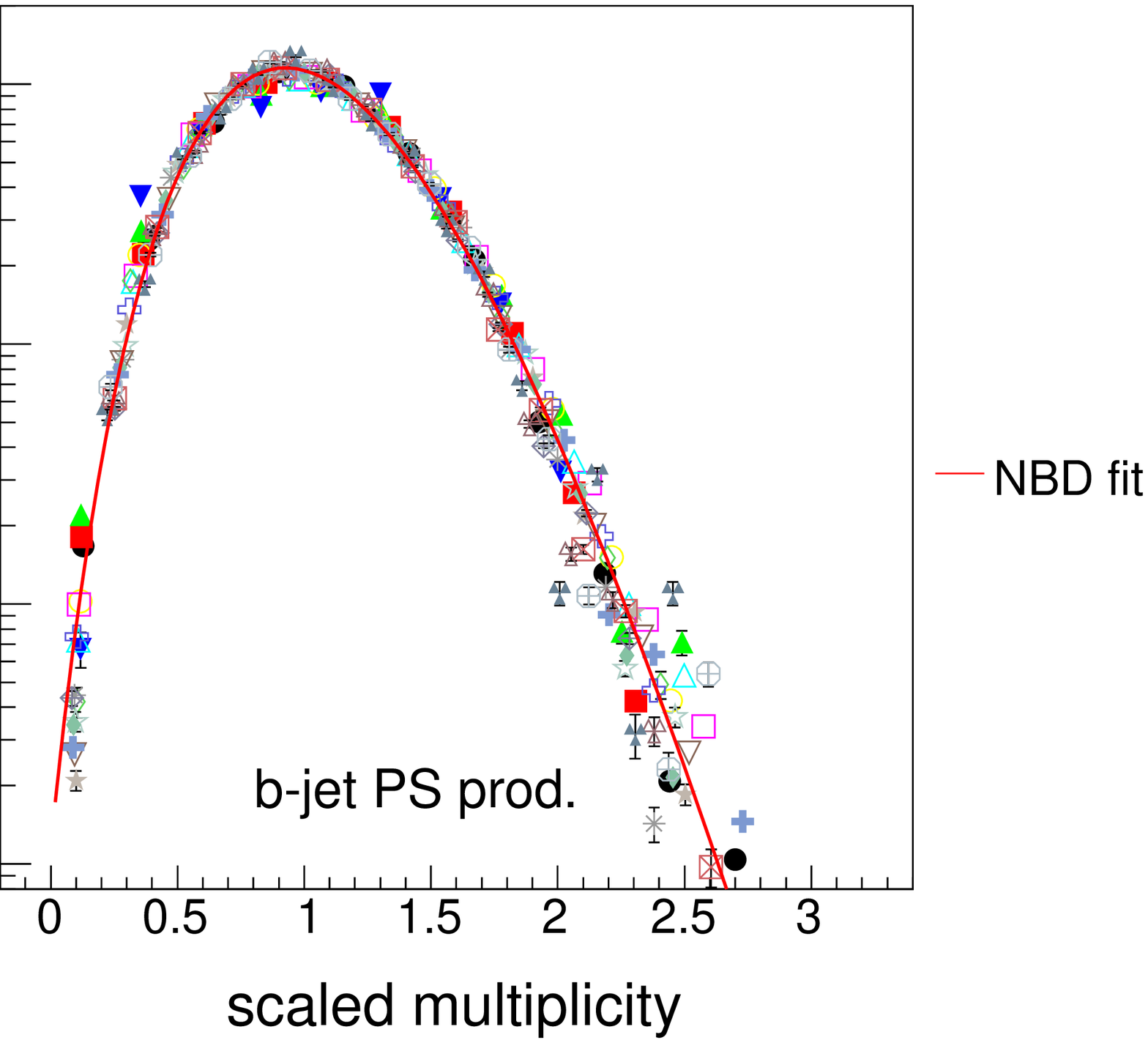}
\caption{Charged-hadron multiplicity distributions with an NBD fit at $|\eta|<1$, for all \pTjet ranges, scaled by the NBD fit means. The four panels from top left to bottom right correspond to inclusive jets, charm and beauty jets from ME flavor creation, and beauty jets from PS production.}
\label{fig:MultDist}
\end{figure}

As a next step we fitted the multiplicity distributions with a negative binomial distribution function in each of the jet transverse momentum ranges,
\begin{equation}
P_N = \frac{\Gamma(Nk+a)}{\Gamma(a)\Gamma(Nk+1)}p^{Nk}(1-p)^a,
\end{equation}
where $a$, $k$ and $p$ are parameters related to the mean and dispersion of the distribution of the multiplicity $N$. In Fig.~\ref{fig:MultDist} we show the multiplicity distributions after all the \pTjet ranges have been scaled on top of each other using the NBD fits. The scaling approximately holds for all four jet samples we investigated. However, for jets containing charm or beauty from flavor creation, the data show minor departures from the NBD fits: the distribution is wider for larger \pTjet, while narrower for smaller \pTjet values.

To quantify the deviations from the scaling behavior, and also to mitigate the effect of fluctuations, we calculated the higher moments of the multiplicity distributions in a similar manner to~\cite{Vertesi:2020utz}. Here the $q^{th}$ moment in a given \pTjet window is defined as

\begin{equation}
\langle N^q\rangle = \sum_{N=1}^{+\infty} P_N N^q ,
\end{equation}

where $P_N$ is the probability distribution corresponding to the event multiplicity $N$. If the scaling is fulfilled and the mean of the distribution scales with a factor $\lambda$, then it is expected that the $q^{th}$ moment scales with $\lambda^q$ as

\begin{equation}
\langle N^q(\pTjet)\rangle = \lambda^q(\pTjet) \langle N^q(p_0)\rangle,
\end{equation}

where $p_0$ is chosen so that the scaling factor is $\lambda(p_0) = 1$. 

\begin{figure}[ht!]
\centering
\includegraphics[width=0.4\columnwidth]{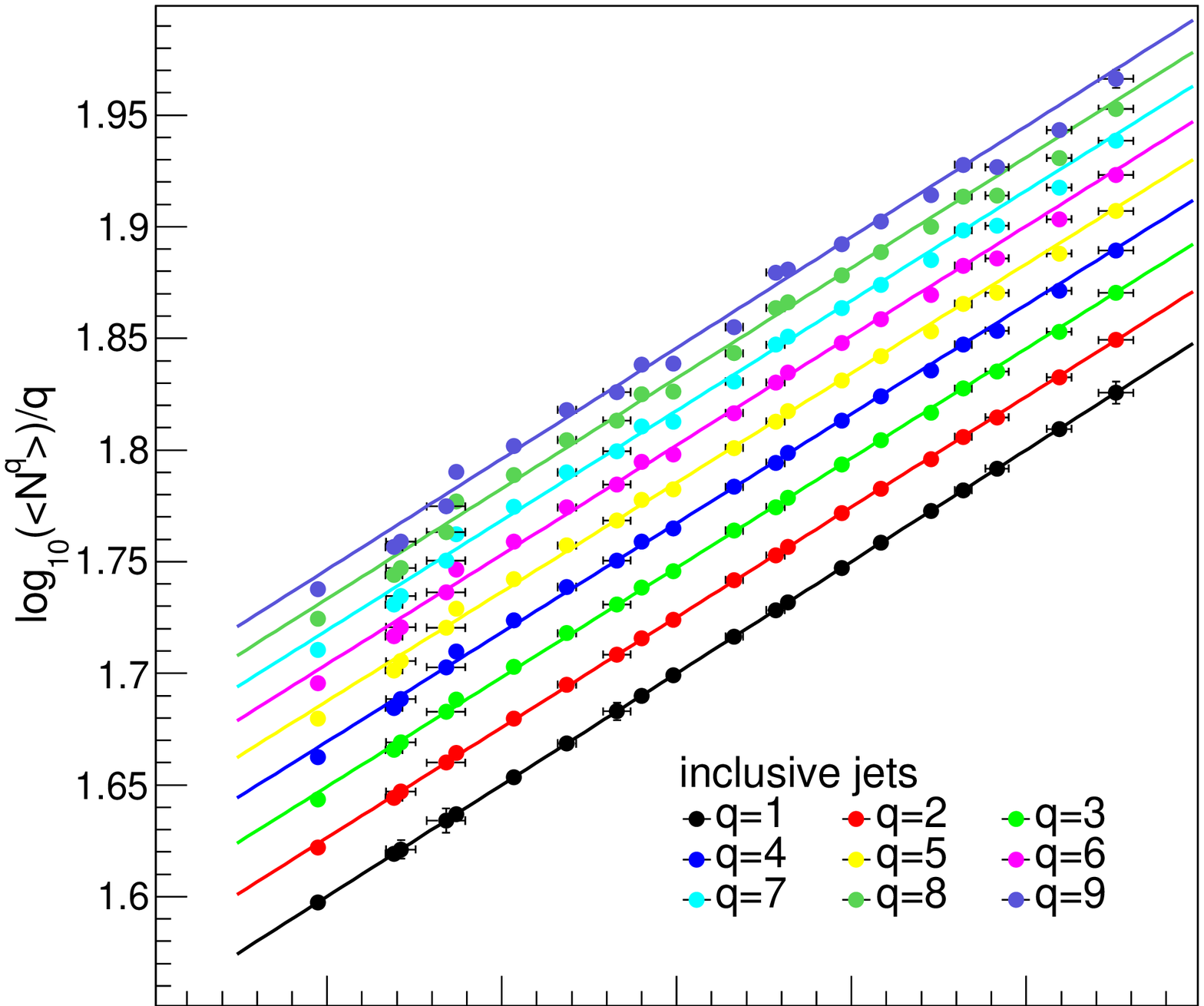}%
\includegraphics[width=0.4\columnwidth]{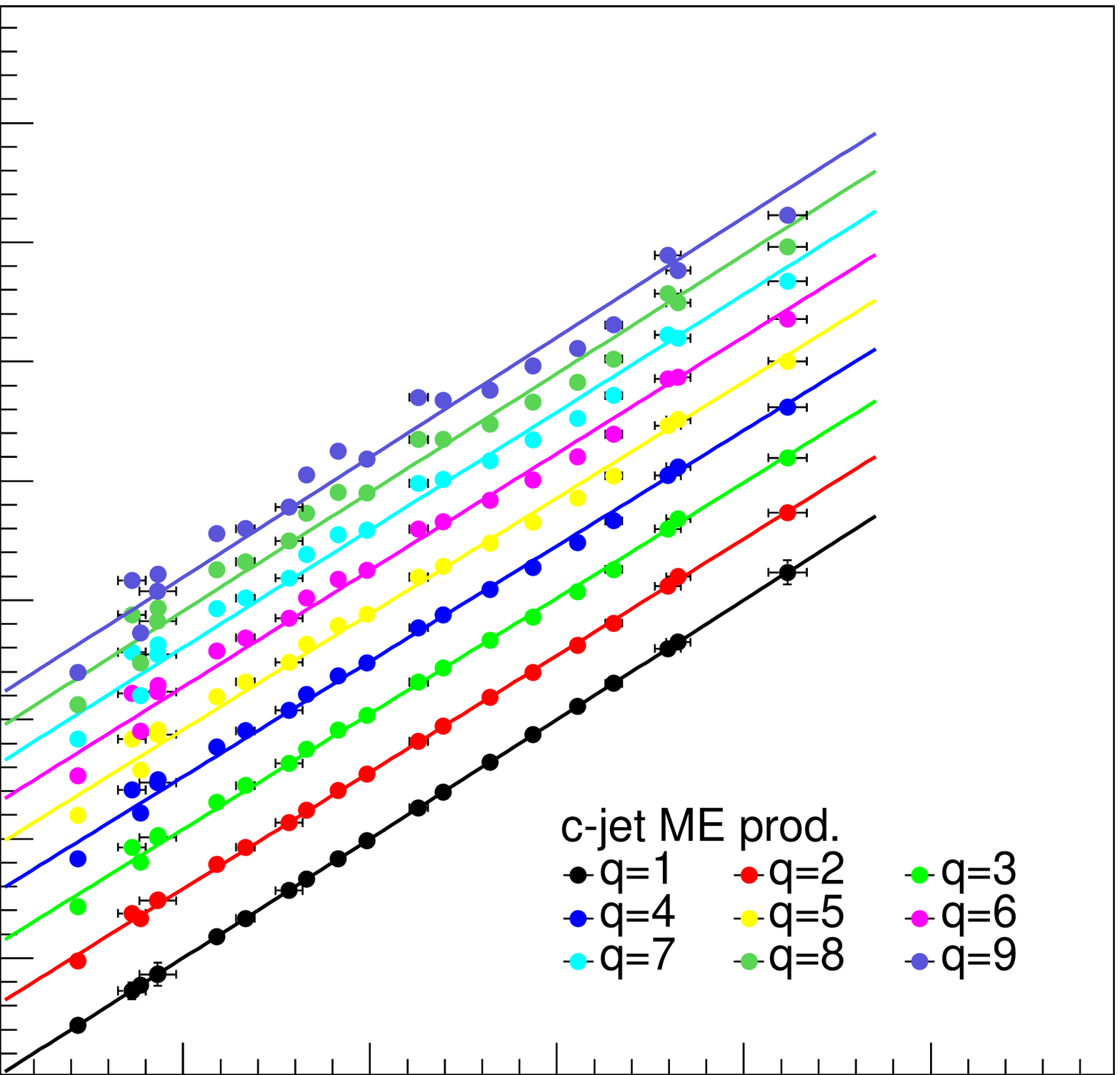}
\includegraphics[width=0.4\columnwidth]{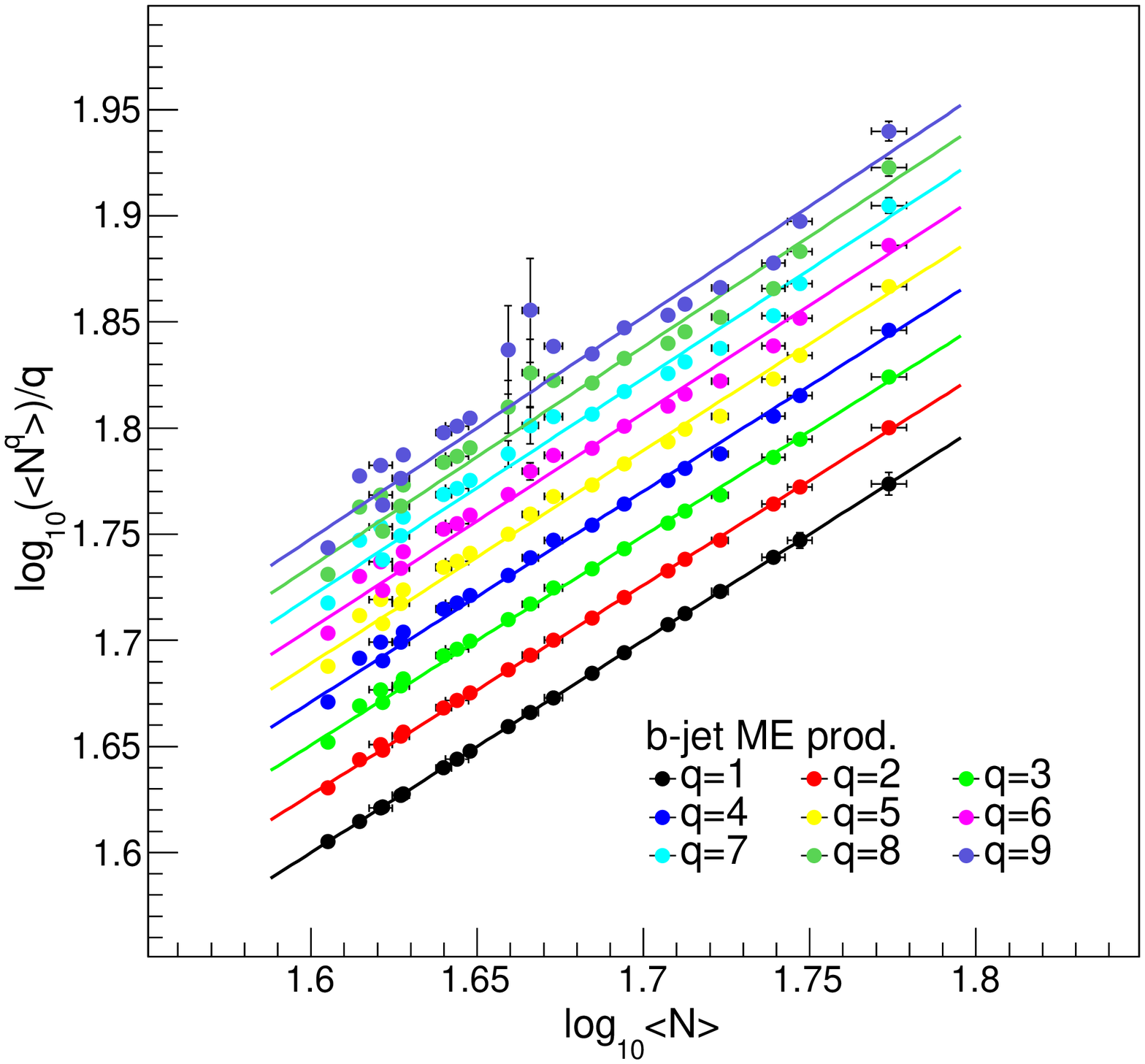}%
\includegraphics[width=0.4\columnwidth]{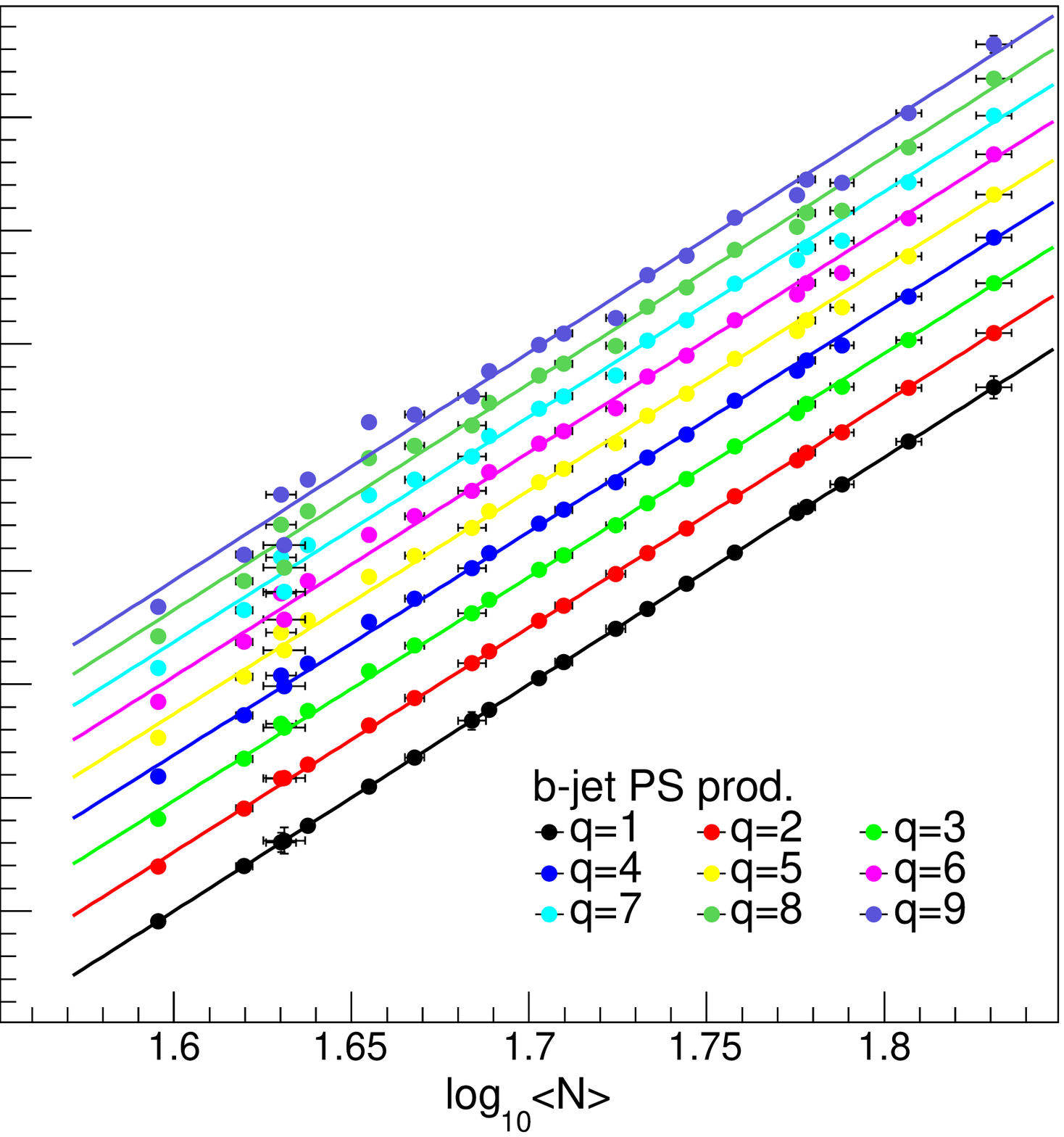}
\caption{The first nine moments of the charged-hadron multiplicity distributions at $|\eta|$<1, in function of the average multiplicity corresponding to each \pTjet range. The four panels are for inclusive jets, charm and beauty jets from ME flavor creation, and beauty jets from PS production. The distributions are normalized by their order $q$ on a log-log scale, and linear fits are applied.}
\label{fig:ScaledMoments}
\end{figure}

In Fig.~\ref{fig:ScaledMoments} we show the first nine moments of the multiplicity distributions divided by their order $q$ in function of the mean charged-hadron multiplicity $\langle N \rangle$ at $|\eta|<1$,
on a log-log scale. The four panels correspond to the four jet categories. The linear fits show a similar trend for all four cases, which means that the scaling is present also for heavy-flavor jets.

\begin{figure}[ht!]
\centering
\includegraphics[width=0.5\columnwidth]{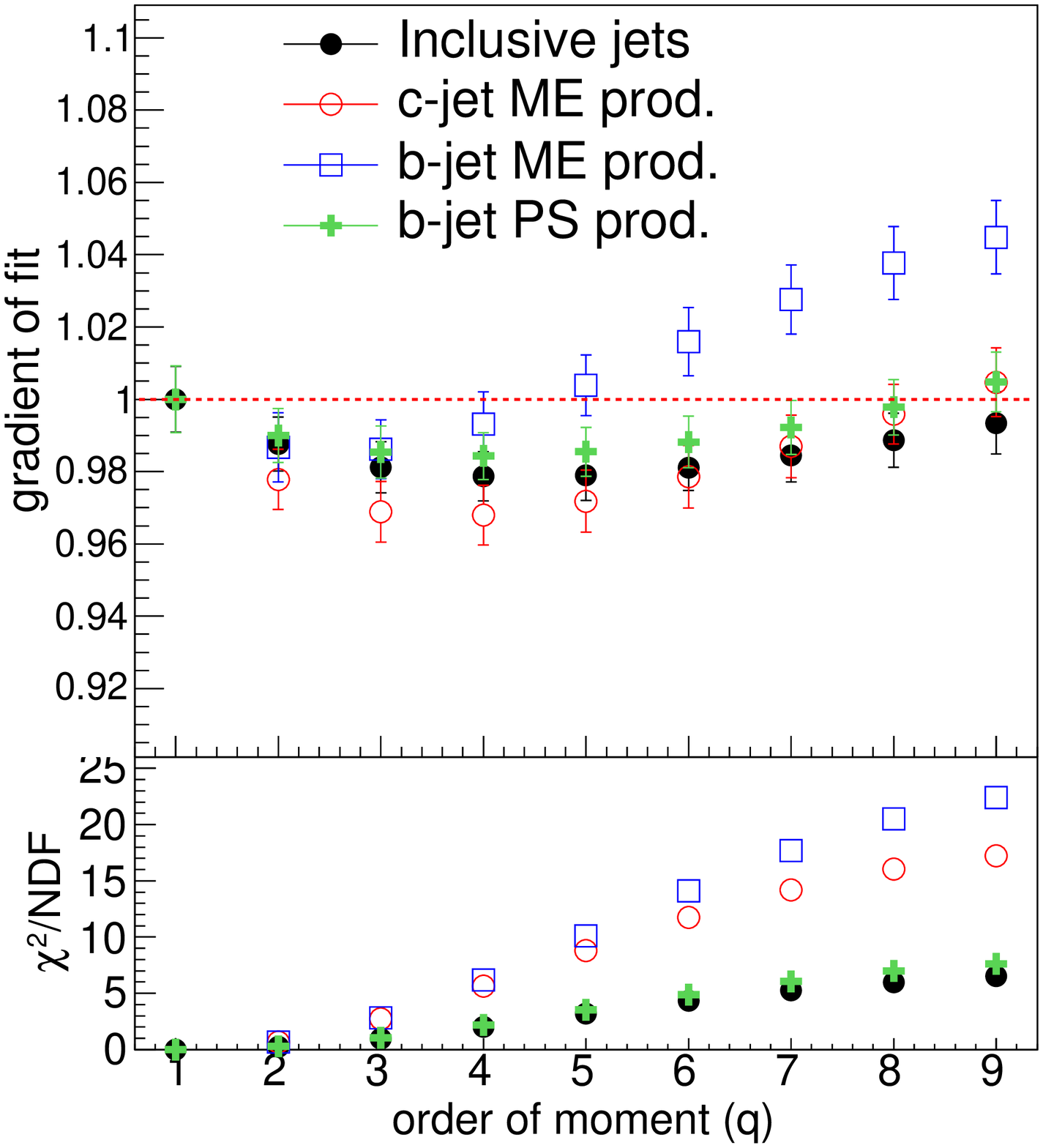}%
\caption{The slope parameters (top panel) and the goodness-of-fit parameters $\chi^2/NDF$ (bottom panel) of the linear fits for the first nine statistical moments of the multiplicity distributions, for charm and beauty jets from ME production and beauty jets from PS production, compared to that for inclusive jets, in function of the order of moments of the multiplicity distributions.}
\label{fig:Gradients}
\end{figure}

Fig.~\ref{fig:Gradients} summarizes the slopes of the fits for the first nine statistical moments, as well as the goodness-of-fit parameter $\chi^2/NDF$. All slopes are around unity within $\approx$5\%. As expected, the goodness of the linear fits is worse for higher moments. The b-jets coming from parton shower processes follow the same trend as the inclusive jets. On the other hand, heavy-flavor jets coming from matrix element production in the simulations correspond to slope parameters that are slightly but significantly different from unity: in case of charm, slopes of the fits for moments $2\leq q \leq 6$ tend to be lower than unity, while in the beauty case the slopes for moments $q \geq 7$ are larger than unity. Furthermore, the goodness of fits for HF ME production tends to be worse than for inclusive jets, $\chi^2/NDF\geq 10$ for any $q\geq 5$. This suggests that the KNO-like scaling originates from the hard parton production, and it is less influenced by the parton shower. The similar patterns of the inclusive jets and the b-jets from PS also indicate that event multiplicities are not driven by flavor-dependent jet fragmentation processes.

%\begin{figure}[H]
%\includegraphics[width=10.5 cm]{Definitions/logo-mdpi}
%\caption{This is a figure. Schemes follow the same formatting. If there are multiple panels, they should be listed as: (\textbf{a}) Description of what is contained in the first panel. (\textbf{b}) Description of what is contained in the second panel. Figures should be placed in the main text near to the first time they are cited. A caption on a single line should be centered.\label{fig1}}
%\end{figure}   

% The MDPI table float is called specialtable
%\begin{specialtable}[H] 
%\caption{This is a table caption. Tables should be placed in the main text near to the first time they are~cited.\label{tab1}}
%%% \tablesize{} %% You can specify the fontsize here, e.g., \tablesize{\footnotesize}. If commented out \small will be used.
%\begin{tabular}{ccc}
%\toprule
%\textbf{Title 1}	& \textbf{Title 2}	& \textbf{Title 3}\\
%\midrule
%Entry 1		& Data			& Data\\
%Entry 2		& Data			& Data\\
%\bottomrule
%\end{tabular}
%\end{specialtable}

%\end{paracol}
% Example of a figure that spans the whole page width (the commands \widefigure and \begin{paracol}{2}, \linenumbers, and\switchcolumn need to be present). The same concept works for tables, too.
%\begin{figure}[H]	
%\widefigure
%\includegraphics[width=15 cm]{Definitions/logo-mdpi}
%\caption{This is a wide figure.\label{fig2}}
%\end{figure}  
%\begin{paracol}{2}
%\linenumbers
%\switchcolumn

\section{Conclusions}
\label{sec:conclusions}

In this manuscript we summarize our results on the scaling properties of heavy-flavor jets from different production processes, and compare them to those on inclusive jets. We used PYTHIA 8 simulations to evaluate the charged-hadron event multiplicities at central pseudorapidity, in function of the charged-particle jet transverse momentum within the range $15<\pTjet<400$ GeV/$c$. We found that the multiplicity distributions satisfy a KNO-like scaling with \pTjet for charm and beauty jets similarly to what has been observed for inclusive jets. We note, however, that multiplicity distributions in events with jets initiated by charm and beauty directly from the leading hard process show some departure from the negative binomial shape, depending on the \pTjet.
Further analysis of the statistical moments of the multiplicity distributions shows that the scaling is fulfilled within $\approx$5\% throughout the full \pTjet range, but the deviations are more significant for leading-order heavy flavor creation, especially in the case of beauty. On the other hand, beauty production from the parton shower tends to deviate less from scaling expectations and follows the inclusive-jet trend within uncertainties. We conclude therefore that the KNO-like scaling originates from the parton level of the early stages of the collision, and not from the later stages of parton shower or jet fragmentation. 

A good description of hadron multiplicity distributions is a basic requirement for models and it is generally fulfilled by the most widely used event generators. However, multiplicities in function of the jet momentum for jets tagged with different flavors can provide means to further validate heavy-flavor production and fragmentation models. Also, while event multiplicity is a good proxy for jet multiplicity in case of jets coming from the leading hard process, this is not necessarily the case for jets that come from secondary hard processes or gluon radiation. An interesting extension of the current work in this direction could therefore be to evaluate the scaling in terms of the jet multiplicity instead of event multiplicity, and to see whether in that case the scaling of heavy flavor jets from the parton shower follows light or heavy jets.
%%%%%%%%%%%%%%%%%%%%%%%%%%%%%%%%%%%%%%%%%%
\vspace{6pt} 

%%%%%%%%%%%%%%%%%%%%%%%%%%%%%%%%%%%%%%%%%%
%% optional
%\supplementary{The following are available online at \linksupplementary{s1}, Figure S1: title, Table S1: title, Video S1: title.}


%%%%%%%%%%%%%%%%%%%%%%%%%%%%%%%%%%%%%%%%%%
\authorcontributions{Conceptualization, R.V.; methodology, R.V. and Z.V.; software, R.V. and Z.V.; validation, V.Z.; formal analysis, Z.V.; investigation, Z.V. and R.V.; resources, R.V.; writing---original draft preparation, Z.V.; writing---review and editing, R.V.; visualization, Z.V.; supervision, R.V.; project administration, R.V.; funding acquisition, R.V. All authors have read and agreed to the published version of the manuscript.}

\funding{This work has been supported by the NKFIH grants OTKA FK131979 and K135515, as well as by the 2019-2.1.11-T\'ET-2019-00078 and 2019-2.1.6-NEMZ\_KI-2019-00011 projects.}

%\institutionalreview{}

%\informedconsent{}

\dataavailability{Data reported here is available from the authors.}

\acknowledgments{The authors acknowledge the computational resources provided by the Wigner GPU Laboratory and the research infrastructure provided by the Eötvös Loránd Research Network (ELKH).}

\conflictsofinterest{The authors declare no conflict of interest.} 

%%%%%%%%%%%%%%%%%%%%%%%%%%%%%%%%%%%%%%%%%%
%\end{paracol}

\reftitle{References}

\bibliography{HF_scaling.bib}

\begin{thebibliography}{999}

\bibitem[Chew and Lim(1985)]{Chew:1985qsa}
Chew, C.K.; Lim, Y.K.
\newblock {Charged Particle Multiplicity Distributions in $e^+ e^-$
  Annihilation and Negative Binomial Distributions}.
\newblock {\em Phys. Lett. B} {\bf 1985}, {\em 163},~257--260.
\newblock
  doi:{\changeurlcolor{black}\href{https://doi.org/10.1016/0370-2693(85)90233-3}{\detokenize{10.1016/0370-2693(85)90233-3}}}.

\bibitem[Giovannini and Van~Hove(1986)]{Giovannini:1985mz}
Giovannini, A.; Van~Hove, L.
\newblock {Negative Binomial Multiplicity Distributions in High-Energy Hadron
  Collisions}.
\newblock {\em Z. Phys. C} {\bf 1986}, {\em 30},~391.
\newblock
  doi:{\changeurlcolor{black}\href{https://doi.org/10.1007/BF01557602}{\detokenize{10.1007/BF01557602}}}.

\bibitem[Acharya \em{et~al.}(2017)Acharya et~al.]{ALICE:2017pcy}
Acharya, S.; others.
\newblock {Charged-particle multiplicity distributions over a wide
  pseudorapidity range in proton-proton collisions at $\sqrt{s}=$ 0.9, 7, and 8
  TeV}.
\newblock {\em Eur. Phys. J. C} {\bf 2017}, {\em 77},~852,
  \href{http://xxx.lanl.gov/abs/1708.01435}{{\normalfont
  [arXiv:hep-ex/1708.01435]}}.
\newblock
  doi:{\changeurlcolor{black}\href{https://doi.org/10.1140/epjc/s10052-017-5412-6}{\detokenize{10.1140/epjc/s10052-017-5412-6}}}.

\bibitem[Koba \em{et~al.}(1972)Koba, Nielsen, and Olesen]{Koba:1972ng}
Koba, Z.; Nielsen, H.B.; Olesen, P.
\newblock {Scaling of multiplicity distributions in high-energy hadron
  collisions}.
\newblock {\em Nucl. Phys. B} {\bf 1972}, {\em 40},~317--334.
\newblock
  doi:{\changeurlcolor{black}\href{https://doi.org/10.1016/0550-3213(72)90551-2}{\detokenize{10.1016/0550-3213(72)90551-2}}}.

\bibitem[Polyakov(1970)]{Polyakov:1970lyy}
Polyakov, A.M.
\newblock {A Similarity hypothesis in the strong interactions. 1. Multiple
  hadron production in e+ e- annihilation}.
\newblock {\em Zh. Eksp. Teor. Fiz.} {\bf 1970}, {\em 59},~542--552.

\bibitem[Alner \em{et~al.}(1986)Alner et~al.]{UA5:1985fid}
Alner, G.J.; others.
\newblock {Scaling Violations in Multiplicity Distributions at 200-GeV and
  900-GeV}.
\newblock {\em Phys. Lett. B} {\bf 1986}, {\em 167},~476--480.
\newblock
  doi:{\changeurlcolor{black}\href{https://doi.org/10.1016/0370-2693(86)91304-3}{\detokenize{10.1016/0370-2693(86)91304-3}}}.

\bibitem[Arneodo \em{et~al.}(1987)Arneodo et~al.]{EuropeanMuon:1987tnv}
Arneodo, M.; others.
\newblock {Comparison of Multiplicity Distributions to the Negative Binomial
  Distribution in Muon - Proton Scattering}.
\newblock {\em Z. Phys. C} {\bf 1987}, {\em 35},~335.
\newblock [Erratum: Z.Phys.C 36, 512 (1987)],
  doi:{\changeurlcolor{black}\href{https://doi.org/10.1007/BF01570769}{\detokenize{10.1007/BF01570769}}}.

\bibitem[Lam and Walton(1984)]{Lam:1983vw}
Lam, C.S.; Walton, M.A.
\newblock {A Proposal for the Origin of {KNO} Scaling}.
\newblock {\em Phys. Lett. B} {\bf 1984}, {\em 140},~246--248.
\newblock
  doi:{\changeurlcolor{black}\href{https://doi.org/10.1016/0370-2693(84)90928-6}{\detokenize{10.1016/0370-2693(84)90928-6}}}.

\bibitem[Kudo(1985)]{Kudo:1985qfd}
Kudo, K.
\newblock {Study of the Violation of {KNO} Scaling and the Validity of Modified
  {KNO} Scaling Based on the Cluster Model}.
\newblock {\em Prog. Theor. Phys.} {\bf 1985}, {\em 74},~1281--1289.
\newblock
  doi:{\changeurlcolor{black}\href{https://doi.org/10.1143/PTP.74.1281}{\detokenize{10.1143/PTP.74.1281}}}.

\bibitem[Burgers \em{et~al.}(1987)Burgers, Hagedorn, and
  Kuvshinov]{Burgers:1987za}
Burgers, G.J.H.; Hagedorn, R.; Kuvshinov, V.
\newblock {Multiplicity distributions in high-energy collisions derived from
  the statistical bootstrap model}.
\newblock {\em Phys. Lett. B} {\bf 1987}, {\em 195},~507.
\newblock
  doi:{\changeurlcolor{black}\href{https://doi.org/10.1016/0370-2693(87)90059-1}{\detokenize{10.1016/0370-2693(87)90059-1}}}.

\bibitem[Hegyi(1997)]{Hegyi:1996wb}
Hegyi, S.
\newblock {Renormalization group approach to multiparticle density
  fluctuations}.
\newblock {\em Phys. Lett. B} {\bf 1997}, {\em 411},~321--325,
  \href{http://xxx.lanl.gov/abs/hep-ph/9612309}{{\normalfont
  [hep-ph/9612309]}}.
\newblock
  doi:{\changeurlcolor{black}\href{https://doi.org/10.1016/S0370-2693(97)00972-6}{\detokenize{10.1016/S0370-2693(97)00972-6}}}.

\bibitem[Matinyan and Walker(1999)]{Matinyan:1998ja}
Matinyan, S.G.; Walker, W.D.
\newblock {Multiplicity distribution and mechanisms of the high-energy hadron
  collisions}.
\newblock {\em Phys. Rev. D} {\bf 1999}, {\em 59},~034022,
  \href{http://xxx.lanl.gov/abs/hep-ph/9801219}{{\normalfont
  [hep-ph/9801219]}}.
\newblock
  doi:{\changeurlcolor{black}\href{https://doi.org/10.1103/PhysRevD.59.034022}{\detokenize{10.1103/PhysRevD.59.034022}}}.

\bibitem[V\'ertesi \em{et~al.}(2021)V\'ertesi, G\'emes, and
  Barnaf\"oldi]{Vertesi:2020utz}
V\'ertesi, R.; G\'emes, A.; Barnaf\"oldi, G.G.
\newblock {Koba-Nielsen-Olesen-like scaling within a jet in proton-proton
  collisions at LHC energies}.
\newblock {\em Phys. Rev. D} {\bf 2021}, {\em 103},~L051503,
  \href{http://xxx.lanl.gov/abs/2012.01132}{{\normalfont
  [arXiv:hep-ph/2012.01132]}}.
\newblock
  doi:{\changeurlcolor{black}\href{https://doi.org/10.1103/PhysRevD.103.L051503}{\detokenize{10.1103/PhysRevD.103.L051503}}}.

\bibitem[Ilten \em{et~al.}(2017)Ilten, Rodd, Thaler, and
  Williams]{Ilten:2017rbd}
Ilten, P.; Rodd, N.L.; Thaler, J.; Williams, M.
\newblock {Disentangling Heavy Flavor at Colliders}.
\newblock {\em Phys. Rev. D} {\bf 2017}, {\em 96},~054019,
  \href{http://xxx.lanl.gov/abs/1702.02947}{{\normalfont
  [arXiv:hep-ph/1702.02947]}}.
\newblock
  doi:{\changeurlcolor{black}\href{https://doi.org/10.1103/PhysRevD.96.054019}{\detokenize{10.1103/PhysRevD.96.054019}}}.

\bibitem[Apolin\'ario \em{et~al.}(2022)Apolin\'ario, Lee, and
  Winn]{Apolinario:2022vzg}
Apolin\'ario, L.; Lee, Y.J.; Winn, M.
\newblock {Heavy quarks and jets as probes of the QGP} {\bf 2022}.
\newblock  \href{http://xxx.lanl.gov/abs/2203.16352}{{\normalfont
  [arXiv:hep-ph/2203.16352]}}.

\bibitem[Dokshitzer \em{et~al.}(1991)Dokshitzer, Khoze, and
  Troian]{Dokshitzer:1991fd}
Dokshitzer, Y.L.; Khoze, V.A.; Troian, S.I.
\newblock {On specific QCD properties of heavy quark fragmentation ('dead
  cone')}.
\newblock {\em J. Phys. G} {\bf 1991}, {\em 17},~1602--1604.
\newblock
  doi:{\changeurlcolor{black}\href{https://doi.org/10.1088/0954-3899/17/10/023}{\detokenize{10.1088/0954-3899/17/10/023}}}.

\bibitem[Acharya \em{et~al.}(2022)Acharya et~al.]{ALICE:2021aqk}
Acharya, S.; others.
\newblock {Direct observation of the dead-cone effect in quantum
  chromodynamics}.
\newblock {\em Nature} {\bf 2022}, {\em 605},~440--446,
  \href{http://xxx.lanl.gov/abs/2106.05713}{{\normalfont
  [arXiv:nucl-ex/2106.05713]}}.
\newblock
  doi:{\changeurlcolor{black}\href{https://doi.org/10.1038/s41586-022-04572-w}{\detokenize{10.1038/s41586-022-04572-w}}}.

\bibitem[Ku\v{c}era(2021)]{Kucera:2021mfd}
Ku\v{c}era, V.
\newblock {Measurements of groomed heavy-flavour jet substructure with ALICE}.
\newblock {\em PoS} {\bf 2021}, {\em HardProbes2020},~149.
\newblock
  doi:{\changeurlcolor{black}\href{https://doi.org/10.22323/1.387.0149}{\detokenize{10.22323/1.387.0149}}}.

\bibitem[Varga \em{et~al.}(2019)Varga, Vértesi, and
  Barnaföldi]{Varga:universe5050132}
Varga, Z.; Vértesi, R.; Barnaföldi, G.G.
\newblock Jet Structure Studies in Small Systems.
\newblock {\em Universe} {\bf 2019}, {\em 5}.
\newblock
  doi:{\changeurlcolor{black}\href{https://doi.org/10.3390/universe5050132}{\detokenize{10.3390/universe5050132}}}.

\bibitem[Sj\"ostrand \em{et~al.}(2015)Sj\"ostrand, Ask, Christiansen, Corke,
  Desai, Ilten, Mrenna, Prestel, Rasmussen, and Skands]{Sjostrand:2014zea}
Sj\"ostrand, T.; Ask, S.; Christiansen, J.R.; Corke, R.; Desai, N.; Ilten, P.;
  Mrenna, S.; Prestel, S.; Rasmussen, C.O.; Skands, P.Z.
\newblock {An introduction to PYTHIA 8.2}.
\newblock {\em Comput. Phys. Commun.} {\bf 2015}, {\em 191},~159--177,
  \href{http://xxx.lanl.gov/abs/1410.3012}{{\normalfont
  [arXiv:hep-ph/1410.3012]}}.
\newblock
  doi:{\changeurlcolor{black}\href{https://doi.org/10.1016/j.cpc.2015.01.024}{\detokenize{10.1016/j.cpc.2015.01.024}}}.

\bibitem[Corke and Sjöstrand(2011)]{Corke_2011}
Corke, R.; Sjöstrand, T.
\newblock Interleaved parton showers and tuning prospects.
\newblock {\em Journal of High Energy Physics} {\bf 2011}, {\em 2011}.
\newblock
  doi:{\changeurlcolor{black}\href{https://doi.org/10.1007/jhep03(2011)032}{\detokenize{10.1007/jhep03(2011)032}}}.

\bibitem[Chatrchyan \em{et~al.}(2013)Chatrchyan et~al.]{CMS:2013ycn}
Chatrchyan, S.; others.
\newblock {Jet and Underlying Event Properties as a Function of
  Charged-Particle Multiplicity in Proton\textendash{}Proton Collisions at
  $\sqrt{s}$ = 7 TeV}.
\newblock {\em Eur. Phys. J. C} {\bf 2013}, {\em 73},~2674,
  \href{http://xxx.lanl.gov/abs/1310.4554}{{\normalfont
  [arXiv:hep-ex/1310.4554]}}.
\newblock
  doi:{\changeurlcolor{black}\href{https://doi.org/10.1140/epjc/s10052-013-2674-5}{\detokenize{10.1140/epjc/s10052-013-2674-5}}}.

\bibitem[Sirunyan \em{et~al.}(2017)Sirunyan et~al.]{CMS:2017eoq}
Sirunyan, A.M.; others.
\newblock {Measurement of charged pion, kaon, and proton production in
  proton-proton collisions at $\sqrt{s}=13$ TeV}.
\newblock {\em Phys. Rev. D} {\bf 2017}, {\em 96},~112003,
  \href{http://xxx.lanl.gov/abs/1706.10194}{{\normalfont
  [arXiv:hep-ex/1706.10194]}}.
\newblock
  doi:{\changeurlcolor{black}\href{https://doi.org/10.1103/PhysRevD.96.112003}{\detokenize{10.1103/PhysRevD.96.112003}}}.

\bibitem[Sjostrand and van Zijl(1987)]{Sjostrand:1987su}
Sjostrand, T.; van Zijl, M.
\newblock {A Multiple Interaction Model for the Event Structure in Hadron
  Collisions}.
\newblock {\em Phys. Rev. D} {\bf 1987}, {\em 36},~2019.
\newblock
  doi:{\changeurlcolor{black}\href{https://doi.org/10.1103/PhysRevD.36.2019}{\detokenize{10.1103/PhysRevD.36.2019}}}.

\bibitem[Sjostrand(1982)]{Sjostrand:1982fn}
Sjostrand, T.
\newblock {The Lund Monte Carlo for Jet Fragmentation}.
\newblock {\em Comput. Phys. Commun.} {\bf 1982}, {\em 27},~243.
\newblock
  doi:{\changeurlcolor{black}\href{https://doi.org/10.1016/0010-4655(82)90175-8}{\detokenize{10.1016/0010-4655(82)90175-8}}}.

\bibitem[Varga \em{et~al.}(2019)Varga, V\'ertesi, and
  G\'abor~Barnaf\"oldi]{Varga:2018isd}
Varga, Z.; V\'ertesi, R.; G\'abor~Barnaf\"oldi, G.
\newblock {Modification of jet structure in high-multiplicity pp collisions due
  to multiple-parton interactions and observing a multiplicity-independent
  characteristic jet size}.
\newblock {\em Adv. High Energy Phys.} {\bf 2019}, {\em 2019},~6731362,
  \href{http://xxx.lanl.gov/abs/1805.03101}{{\normalfont
  [arXiv:hep-ph/1805.03101]}}.
\newblock
  doi:{\changeurlcolor{black}\href{https://doi.org/10.1155/2019/6731362}{\detokenize{10.1155/2019/6731362}}}.

\bibitem[Strong(2015)]{Giles:2015gluon}
Strong, G.C.
\newblock {Gluon splitting to b-quark pairs in proton-proton collisions at
  $\sqrt{s}=8$ TeV with ATLAS}.
\newblock {\em MSc Thesis} {\bf 2015}.

\bibitem[Cacciari \em{et~al.}(2008)Cacciari, Salam, and Soyez]{Cacciari2008}
Cacciari, M.; Salam, G.P.; Soyez, G. {\bf 2008}.
\newblock {\em 2008},~063--063.
\newblock
  doi:{\changeurlcolor{black}\href{https://doi.org/10.1088/1126-6708/2008/04/063}{\detokenize{10.1088/1126-6708/2008/04/063}}}.

\bibitem[Chatrchyan \em{et~al.}(2013)Chatrchyan et~al.]{CMS:2012feb}
Chatrchyan, S.; others.
\newblock {Identification of b-Quark Jets with the CMS Experiment}.
\newblock {\em JINST} {\bf 2013}, {\em 8},~P04013,
  \href{http://xxx.lanl.gov/abs/1211.4462}{{\normalfont
  [arXiv:hep-ex/1211.4462]}}.
\newblock
  doi:{\changeurlcolor{black}\href{https://doi.org/10.1088/1748-0221/8/04/P04013}{\detokenize{10.1088/1748-0221/8/04/P04013}}}.

\bibitem[Acharya \em{et~al.}(2022)Acharya et~al.]{ALICE:2021wct}
Acharya, S.; others.
\newblock {Measurement of inclusive charged-particle b-jet production in pp and
  p-Pb collisions at $ \sqrt{s_{\mathrm{NN}}} $ = 5.02 TeV}.
\newblock {\em JHEP} {\bf 2022}, {\em 01},~178,
  \href{http://xxx.lanl.gov/abs/2110.06104}{{\normalfont
  [arXiv:nucl-ex/2110.06104]}}.
\newblock
  doi:{\changeurlcolor{black}\href{https://doi.org/10.1007/JHEP01(2022)178}{\detokenize{10.1007/JHEP01(2022)178}}}.

\end{thebibliography}

\end{document}